# Studies of dc current-driven switching in Py/N/Py magnetic nanopillars

S. Urazhdin, W.P.Pratt Jr., and J. Bass
*Department of Physics and Astronomy, Center for Fundamental Materials Research, and Center for Sensor Materials, Michigan State University, East Lansing, MI 48824-2320, USA.*

We provide new data on current-driven switching in Permalloy (Py = $Ni_{84}Fe_{16}$) based nanopillars at room temperature and 4.2K that confirm and extend previously published work. We present data for both uncoupled and antiferromagnetically coupled samples. The latter confirm prior results for Co/Cu/Co. We show that inserting sufficiently strong spin-flip-scattering into the Cu layer eliminates hysteretic current-driven switching. This result may have ramifications for understanding current-driven switching.

A fair amount of data and theory [1] has been produced since the first observations [2-4] of predicted [5,6] current-driven magnetic excitations and switching in magnetic nanopillars. While the relatively simple original models qualitatively explain most of these data, more detailed studies and deeper understanding of the relevant physics are still needed to produce devices with desired current-switching properties. Technological interest lies in the possibility of using current-driven switching for magnetic memory writing, and learning how to control noise and switching in potential current-perpendicular-to-plane read heads at large currents. In this paper we present three new sets of data on current-driven switching in Py/Cu/Py (Py = Permalloy = $Ni_{84}Fe_{16}$) multilayers, that extend our earlier studies. The small crystalline anisotropy and magnetoelastic coefficients of Py, allow stable, reproducible data at both room temperature (295K) and 4.2K.

Our samples are nanofabricated Py/Cu/Py trilayer nanopillars with approximately elliptically shaped top Py layers. This shape of the top Py layer minimizes the nonuniformity of its magnetization and usually gives single step monodomain switching. In the first and third studies to be described, we minimized dipolar coupling between the two Py layers by leaving the lower and thicker (pinned) Py layer extended in lateral dimensions (~ microns), and patterning only the upper, thinner (free) Py layer and part of the Cu spacer. The current I needed to affect the magnetization state of the bottom Py layer is then expected to be much larger than that for reversing the moment of the top Py nanopillar. Use of a 6-10 nm thick Cu spacer should give negligible exchange coupling between the two Py layers. Our nanopillars are small enough (~ 70 x 130 nm) so that exchange-related current-driven switching should dominate over self-Oersted-field effects. The second study required antiferromagnetic (AF) coupling between the two Py layers. We achieved such coupling by patterning completely through the Cu layer and partway into the bottom (thick) Py layer. In the third study, on uncoupled samples, we replaced the Cu spacer with a Cu/Cu(Pt)/Cu sandwich (Cu(Pt) = $Cu_{94}Pt_6$), in which the Cu(Pt) thickness was 12 nm and the equal Cu thicknesses were 1.5 nm. Spin-flip scattering in the Cu(Pt) layer reduced the magnetoresistance, and also led to changes in the current-switching, as described below. In all three studies, positive current flows from the pinned to the free Py layer, and H is applied in the layer plane and along the magnetic easy axis of the nanopillar. We measure the dynamic resistance, dV/dI, using four probes and lock-in detection, superimposing a small ac current (~20-40 µA at 8 kHz) onto the dc current I.

Fig. 1 compares data on current-driven switching for a magnetically uncoupled Py(30)/Cu(10)/Py(3.5) trilayer—thicknesses in nm—at 295K (a,c,d) and 4.2K (b,d,f). The main features of the data are consistent with those we published previously for a different Py sample [7]. The magnetization switching gives changes in the dV/dI through the current-perpendicular (CPP) magnetoresistance (MR). Figs. 1a (295K) and 1b (4.2K) show that the magnetization switching driven by the field H at I = 0 occurs in a single sharp step, between the high field, low resistance state of parallel (P) orientation of the Py layer magnetizations and the low field, high resistance state of anti-parallel (AP) orientation. Figs. 1c (295K) and 1d (4.2K) show that the current driven switching at fixed H, is hysteretic with sharp steps at small H, but becomes reversible (indicated by a much higher peak) at larger H. Figs. (e) (295K) and (f) (4.2K) show the variations of the switching current $I_s$ with H, at H>0. The hysteretic switching is symmetric in H, while the reversible switching is more asymmetric, as expected due to the Oersted field for larger I,H (not shown). The quantitative reduction in switching field from prior data at 295K [7] is due to nearly superparamagnetic behavior of the Py(3.5) nanopillar at 295K. Due to thermal activation, the switching field $H_s$ (I=0), $I_s$(H=0), and thus also the bistable region area (Fig.1(e)), are significantly smaller at 295 K than at 4.2 K.

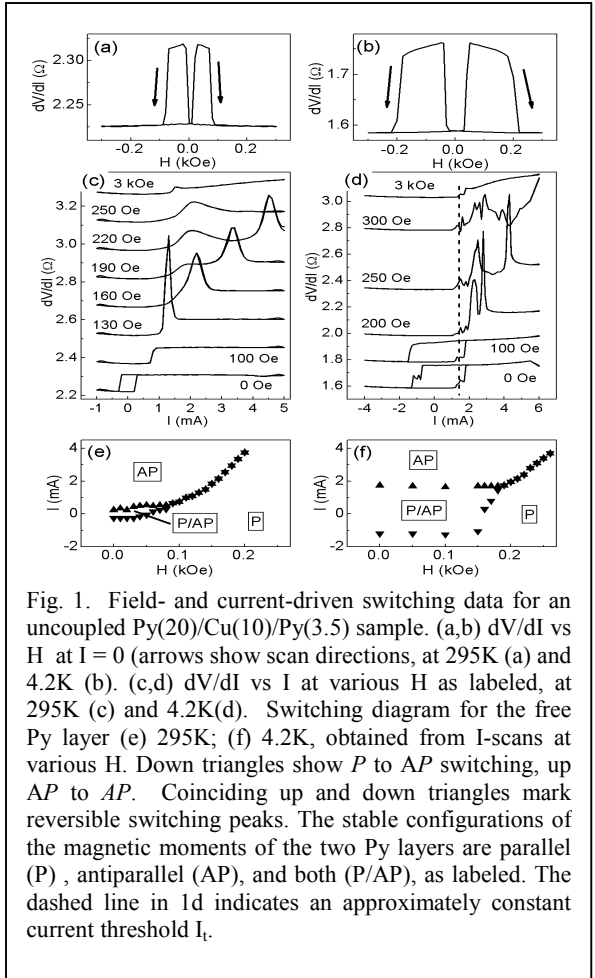

Fig. 1. Field- and current-driven switching data for an uncoupled Py(20)/Cu(10)/Py(3.5) sample. (a,b) dV/dI vs H at I = 0 (arrows show scan directions, at 295K (a) and 4.2K (b). (c,d) dV/dI vs I at various H as labeled, at 295K (c) and 4.2K(d). Switching diagram for the free Py layer (e) 295K; (f) 4.2K, obtained from I-scans at various H. Down triangles show $P$ to $AP$ switching, up $AP$ to $AP$. Coinciding up and down triangles mark reversible switching peaks. The stable configurations of the magnetic moments of the two Py layers are parallel (P), antiparallel (AP), and both (P/AP), as labeled. The dashed line in 1d indicates an approximately constant current threshold $I_t$.

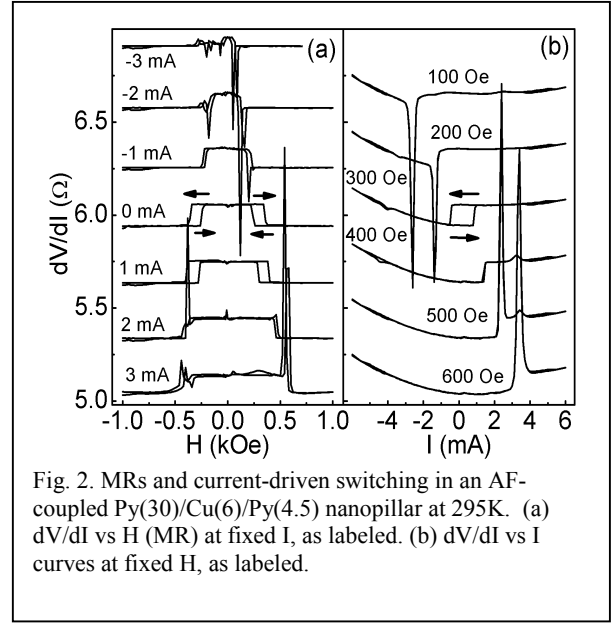

Fig. 2. MRs and current-driven switching in an AF-coupled Py(30)/Cu(6)/Py(4.5) nanopillar at 295K. (a) dV/dI vs H (MR) at fixed I, as labeled. (b) dV/dI vs I curves at fixed H, as labeled.

In ref. [7] we identified a threshold current for large amplitude magnetic excitations (which we labeled by $I_t$) in the reversible region, above which dV/dI increased approximately linearly with I in a standard size nanopillar, or at 4.2 K displayed a series of peaks in a smaller size one. Fig. 1d shows a similar series of peaks at 4.2 K in this new sample, which become smeared out into a 'hump' at 295K (Fig.1c). In [7] we showed that $I_t \approx I_s$(H=0) at 4.2 K, and varied very slowly with H. Fig.1d shows that, at 4.2 K, $I_t$ is also almost independent of H, but in this sample slightly smaller than $I_s$. In the hysteretic region at small H, the current-switching curves in Fig. 1d now contain multiple steps. More careful examination shows that only the largest step at the largest I is due to magnetization switching, while the smaller steps in dV/dI at smaller I are insensitive to H, and are due to magnetic excitations within the P state (i.e., not involving partial reversal of the layer magnetization). Fig. 4c shows that at 295 K, $I_t \gg I_s$(H=0). The correlation $I_t \approx I_s$(H=0) shown in [7] for 4.2 K is thus somewhat violated at 295 K due to thermal activation. This correlation can also be violated in magnetically coupled samples (see [8,9] and below).

In refs. [8,9] we showed differences in behavior of uncoupled and antiferromagnetically (AF) coupled Co/Cu/Co nanopillars at 4.2K. The uncoupled Co/Cu/Co sample behaved similarly to the uncoupled Py/Cu/Py sample of Fig.1, except for a less pronounced threshold behavior. In contrast, the AF-coupled sample showed reversible switching at small H, which changed to hysteretic switching at intermediate H, and then back to reversible switching again at large H. Fig. 2a shows dV/dI vs H at fixed I, and Fig. 2b shows dV/dI vs I at fixed H for an AF-coupled Py(30)/Cu(6)/Py(4.5) sample at T= 295K. The dV/dI vs I data behave very similarly to those for Co/Cu/Co [8,9]. Together, these data yield the switching diagram shown in Fig. 3, in which the hysteretic behavior is centered about the dipole fiel d $H_d \approx 320$ Oe. The symmetry with respect to the reversal of both I and (H-320 Oe) is indicative of the similar effect of I>0 in the P state, and I<0 in the AP state. This symmetry is not present in uncoupled samples, because finite H always biases the stability towards the P state.

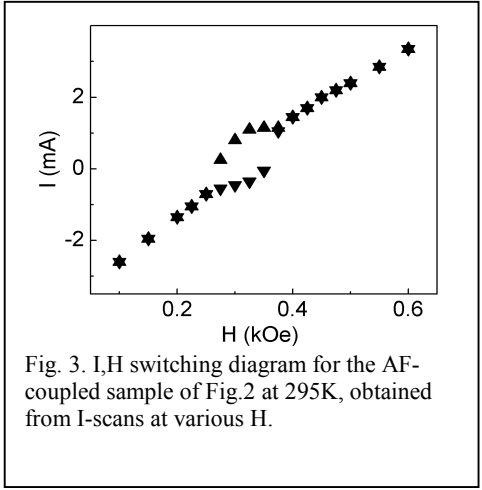

Fig. 3. I,H switching diagram for the AF-coupled sample of Fig.2 at 295K, obtained from I-scans at various H.

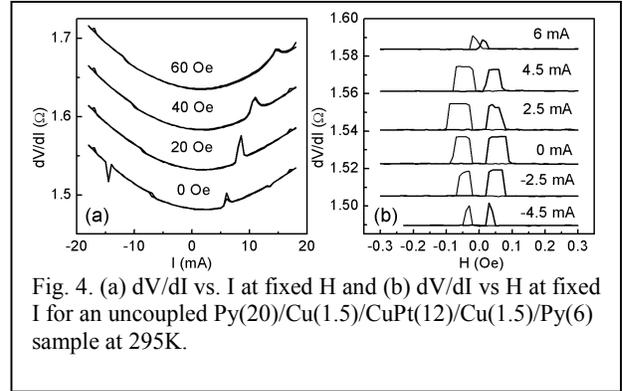

Fig. 4. (a) dV/dI vs. I at fixed H and (b) dV/dI vs H at fixed I for an uncoupled Py(20)/Cu(1.5)/CuPt(12)/Cu(1.5)/Py(6) sample at 295K.

In ref. [10], we showed that when the Cu spacer between the Py layers is replaced by a Cu/Cu(Pt)(d)/Cu sandwich, where (d) is the thickness of the Cu(Pt) layer, the resistance change, $\Delta R = (dV/dI)_{AP} - (dV/dI)_P$, decreased due to spin-flip scattering in the CuPt layer as $\Delta R(d) \approx \Delta R(0)\exp[-d/l_{sf}]$, where $l_{sf} \approx 6$ nm is the spin-diffusion length in CuPt at 295K [10,11]. Simultaneously, $I_s$ increased, so that the product $(\Delta R)I_s$ remained essentially unchanged. Reproducible hysteretic current-driven switching persisted up to d = 8 nm. In contrast, as illustrated in Fig. 4a for H = 0 to 60 Oe, none of our eight samples with d = 12 nm showed reproducible hysteretic current-driven switching. Instead, at small H these samples exhibited reversible switching peaks or steps in dV/dI (upward peaks or upward steps correspond to increases in the resistance V/I), that moved rapidly to higher I as H increased. In several independent samples, the steps in dV/dI (see, e.g., those in Fig. 4a associated with the peaks mostly at positive I, or those by themselves at negative I) were always noticeably smaller than the MR switching steps (as in Fig. 4b). We thus describe such current-driven switching as a transition between the P state and an inhomogeneous magnetization state of the free Py-layer, likely affected by the current-induced Oersted field. We note that the upward step at H = 0, I = -7 mA, indicating switching from P to the inhomogeneous state, is approximately symmetric with the upward peak at I = +6 ma. This upward step is followed by a downward peak at I = -14.5 mA, indicating a partial reverse transition. In some of the other samples with d = 12 nm, the switching peaks equivalent to those for +I in Fig. 4a appeared only at I < 0, or for both directions of I. Averaging over 8 samples, at H = 0 the peaks appeared at $|I| \approx 10$ mA.

In contrast to the peaks in Fig. 4a, Fig. 4b (I = 0 curve) shows that the sample with d = 12 nm exhibits field-driven switching similar in form to samples with smaller d (see, e.g. Fig. 1a,b). However, $H_s$ decreases as $|I|$ increases for both I>0 and I<0. This behavior, shown in the switching diagram of Fig. 5, is qualitatively different from that of uncoupled Co/Cu/Co and Py/Cu/Py samples (Fig.1e,f and [7,8]), where $H_s$ decreases for I<0, but increases for I > 0. The switching diagram of Fig. 5 explains the absence of hysteretic current-switching in Fig. 4a. When I is reduced from a large positive value at small |H|, the system first switches reversibly from the inhomogeneous state to the P state (as I crosses the line marked by open symbols). When I is reduced further, the system goes through the P/AP bistable region into the region where only the P state is stable, without changing its magnetization state. To achieve the AP state in the bistable region, it is necessary to flip the bottom Py layer by varying H.

The results of Figs. 4 and 5 may be important for understanding the current-driven behavior of a single magnetic layer [3,12]. Placing a thick CuPt layer in the spacer between two Py layers lets us approach a single-layer regime, while still retaining the small MR needed to determine the magnetic orientation of the nanopillar. The approximate symmetry of the effects of positive and negative currents (Fig. 5) shows that the switching of the nanopillar is not significantly affected by the bottom Py layer. We propose two alternative explanations for the behaviors of Figs. 4,5, both of which assume that the current arriving at the free layer from the pinned layer is essentially unpolarized. 1) The dominant effect of such an unpolarized current might be due to its Oersted field and/or to Joule heating. Either phenomenon could reduce $H_s$, independent of the current direction. 2) Alternatively, an unpolarized current can spontaneously generate magnetic excitations [13], at a rate that is independent of both the current direction and the orientation of the magnetizations of the nanopillar. Such excitations could be described in terms of an increase of the effective magnetic temperature $T_m$ [7,14]. The decrease of $H_s$ with |I| would then be due to thermally activated switching with $T_m(I) > T_{ph}$, the ambient temperature. More detailed experiments and modeling are needed to distinguish between alternatives 1) and 2).

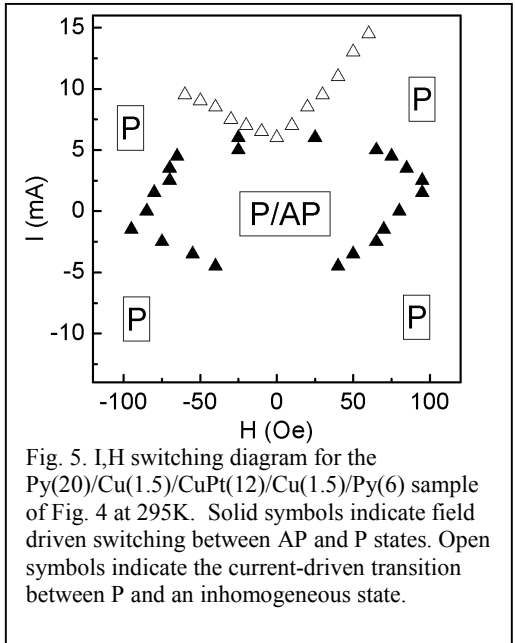

Fig. 5. I,H switching diagram for the Py(20)/Cu(1.5)/CuPt(12)/Cu(1.5)/Py(6) sample of Fig. 4 at 295K. Solid symbols indicate field driven switching between AP and P states. Open symbols indicate the current-driven transition between P and an inhomogeneous state.

To summarize, in this paper we have described three new experiments involving Py/Cu/Py nanopillars. In the first we confirmed several of the behaviors described in a prior study of magnetically uncoupled Py/Cu/Py nanopillars [7], and extended to 295K a comparison between the threshold current $I_t$ for excitations in the reversible regime and the switching current $I_s$ in the hysteretic regime. In the second, we confirmed that antiferromagnetically (AF) coupled Py/Cu/Py nanopillars display variations of switching behavior with current I similar to those previously reported for AF coupled Co/Cu/Co nanopillars. In the third, we showed that increasing the spin-flipping within the Cu layer sufficiently, by inserting a 12 nm thick layer of a strong spin-flipping alloy Cu(Pt), completely eliminates hysteretic current-driven switching, and changes the I,H switching diagram, so that the switching field, $H_s$, decreases with increasing magnitude of the current I, independent of the sign of the current. We suggest two possible explanations for the results of this third experiment, which might have significant implications for the theory of current-driven switching.

Acknowledgments: The authors thank Norman Birge for helpful discussions. This work was partially supported by the MSU CSM, CFMR, MSU Keck Microfabrication Laboratory, NSF grants DMR-98-09688, 02-02476, and NSF-EU 00-98803, plus Seagate Technology.